\documentstyle[12pt,epsf]{article}
\catcode`\@=11

\newcommand{\dr}{\mbox{\footnotesize$\overline{\rm DR}$~}}

\newcommand{\blog}{\overline{\ln}}
\newcommand{\gsim}{\lower.7ex\hbox{$\;\stackrel{\textstyle>}{\sim}\;$}}
\newcommand{\lsim}{\lower.7ex\hbox{$\;\stackrel{\textstyle<}{\sim}\;$}}

\begin{document}
\begin{titlepage}
\samepage{
\setcounter{page}{1}
\rightline{MADPH-98-1072}
\rightline{hep-ph/9808299}
\rightline{August, 1998}
\vfill
\vfill
\begin{center}
{\bf TWO-LOOP EFFECTIVE POTENTIAL CALCULATION OF 
THE LIGHTEST $CP$-EVEN HIGGS-BOSON MASS IN THE MSSM}
\vfill
 {REN-JIE ZHANG\footnote{\tt rjzhang@pheno.physics.wisc.edu}\\}
\vspace{.25in}
{\it Department of Physics\\
     University of Wisconsin\\
     1150 University Avenue\\
     Madison, Wisconsin 53706 USA\\}
\end{center}
\vfill
\vfill
\begin{abstract}
{\rm 
We calculate a two-loop effective potential
to the order of ${\cal O}(\lambda_t^2\alpha_s)$ in the MSSM.
We then study the corresponding two-loop corrections to the
$CP$-even Higgs-boson mass for arbitrary $\tan\beta$ 
and left-right top-squark mixings. 
We find that the lightest
Higgs-boson mass is changed by at most a few GeV. We also show the
improved scale dependence and compare to previous two-loop analyses.}
\end{abstract}
\vfill}
\end{titlepage}

\section{Introduction}

In the minimal supersymmetric standard model (MSSM), the Higgs
sector is composed of three neutral (two $CP$-even, one 
$CP$-odd) and two charged scalar bosons. 
An important fact of the model is that
the quartic self-coupling of the lightest $CP$-even Higgs boson $h$
is not a free parameter, but 
related to the standard model gauge couplings $g$ and $g'$;
as a result, the Higgs boson has an 
upper bounded tree-level mass, $m_h \leq M_Z$. 
This limit, however, is violated when the
one-loop radiative corrections are included.

The well-known dominant one-loop radiative correction comes from
the incomplete cancellation of the virtual top-quark
and top-squark loops \cite{1-loop}, it approximately has the size  
$\Delta m_h^2\simeq {3\lambda_t^2 m_t^2\over 4\pi^2}
\log({m^2_{\tilde t}/m^2_t})$. Taking 
$m_{\tilde t}= 100-1000$ GeV,
one finds a large correction, $\Delta m_h\simeq
{\rm a\ few}-50$ GeV, due to the 
relatively large top-quark mass $m_t^{\rm pole}=175$ GeV.
The one-loop Higgs-boson mass sensitively depends
on the top-quark mass, and generally varies with the change of
renormalization scale.
So it remains a quite important problem to study the magnitude 
of two-loop radiative corrections and the scale 
dependence of the Higgs-boson mass after these corrections.

There are basically two approaches to calculate the two-loop 
radiative corrections.  In the renormalization group 
equation (RGE)-improved effective potential approach
\cite{2-loop, Casas, CQW}, 
one uses the one-loop effective potential, together
with two-loop RGEs, then all leading-order 
and next-to-leading-order corrections
can be calculated. The finite one-loop threshold
corrections arising from the decouplings of the 
heavy top-squarks have also been 
included, both for the small and large left-right
top-squark mixings.
It is further observed
that by judiciously setting the renormalization scale,
one can use the one-loop renormalized Higgs-boson mass as a good
approximation to the full two-loop results \cite{Casas, Haber}. 

The second approach involves a two-loop effective potential.
In the special case of $\tan\beta \rightarrow \infty$ 
and no left-right mixing in the top-squark sector,
Hempfling and Hoang have calculated the upper bound to the
two-loop Higgs-boson mass \cite{HH}. Their results qualitatively 
agree with the previous approach. Recently, two-loop corrections to $m_h$  
have also been computed by an explicit diagrammatic method \cite{hollik}.

It is the purpose of this paper to generalize the two-loop
effective potential calculation of \cite{HH} to the case of
arbitrary $\tan\beta$ and left-right top-squark 
mixings. The effective potential method has an advantage over other
approaches because it is simple and does 
not require complicate programming.
In this paper we will study the improved scale dependence 
of the Higgs-boson mass $m_h$
and the size of two-loop corrections. 
We will also compare our results with previous two-loop
calculations \cite{CQW, hollik}. 
 
The rest of the paper is organized as follows: We first present a general
formalism for calculating the $CP$-even Higgs-boson mass from the effective
potential in Section 2, and compute the two-loop effective potential to the
order of ${\cal O}(\lambda^2_t\alpha_s)$. We next show
the results of our numerical analyses in Section 3; we find good agreements 
with previous two-loop calculations. Finally we conclude in Section 4.
For completeness, some functions
which appear in the two-loop calculation 
are given in the Appendix. 

\section{Effective Potential and the $CP$-even Higgs-boson Masses}

We start our analysis with the tree-level 
potential of the MSSM\footnote{We 
work in the modified $\dr$-scheme of Ref. \cite{dr}.},
\begin{eqnarray}
V_{\rm tree}&=&(m^2_{H_1}+\mu^2)|H_1|^2 + (m^2_{H_2}+\mu^2)|H_2|^2 
+\mu B (H_1 H_2 +{\rm H.c.})\nonumber\\ 
&&\qquad\qquad\qquad+ {g^2+g'^2\over 8}(|H_1|^2-|H_2|^2)^2
+ {g^2\over 2}|H_1^\dagger H_2|^2,\label{eq:V0}
\end{eqnarray}
where $g, g'$ are the $SU(2)$ and $U(1)_Y$ gauge couplings, 
$m_{H_1}, m_{H_2}$ and $B$ 
are the soft-breaking Higgs-sector mass parameters,
and $\mu$ is the supersymmetric Higgs-boson mass parameter 
(the $\mu$-parameter).

We express $H_1$ and $H_2$ in terms of their component fields, 
\begin{equation}
H_1=\left(\begin{array}{c}
{(S_1 + i P_1)/{\sqrt 2}}\\[2mm]
 H_1^-
\end{array}\right), \qquad
H_2=\left(\begin{array}{c}
H_2^+\\[2mm]
{(S_2 + i P_2)/{\sqrt 2}}
\end{array}\right),
\end{equation}
so the tree-level potential can be rewritten as a function of 
the $CP$-even fields, $S_1$ and $S_2$. In general,
the all-loop effective potential is a function of $S_1$ and
$S_2$, which are usually known as classical fields. 
 
The technique for calculating a higher loop effective potential
was developed long ago by Jackiw \cite{jackiw}. First,
the Higgs fields are expanded around the classical fields,
in terms of which all the relevant 
particle masses and couplings are determined. 
One then calculates the higher loop effective potential by computing the
corresponding zero-point function Feynman diagrams (bubble diagrams).

To be more specific,
to the two-loop order that we consider in this paper, 
we write the effective potential as
\begin{equation}
V(S_1, S_2)\ =\ V_0 + V_{\rm tree}(S_1,S_2) 
+ V_{\rm 1loop}(S_1,S_2) + V_{\rm 2loop}(S_1,S_2)\ ,
\end{equation} 
where $V_0$ is a field-independent vacuum-energy term, and 
$V_{\rm tree}$, $V_{\rm 1loop}$ and $V_{\rm 2loop}$ are
the tree-level, one- and two-loop contributions respectively.
$V_0$ is necessary for the renormalization group 
invariance of the effective potential \cite{kastening}.

The one-loop effective potential in the $\dr$  
scheme is well-known. It can be easily obtained by calculating 
the one-loop bubble diagrams with all kinds of (s)particles 
running  in the loop. We find that in the Landau gauge
\begin{eqnarray}
V_{\rm 1loop}(S_1, S_2)
 &=& \sum_f\sum_{i=1}^2 N_c^f G({\tilde f_i}) - 2 \sum_{f}
N_c^f G(f) + 3 G(W) + {3\over 2} G(Z) \nonumber\\ 
&+ &  {1\over 2}\biggl[G(H) + G(h) + G(A) + G(G)\biggr]
+ G({H^+}) + G({G^+}) \nonumber\\
&- &  2 \sum_{i=1}^2 G({{\tilde\chi}^+_i}) 
- \sum_{i=1}^4 G({{\tilde\chi}^0_i})\ ,
\label{1lep} 
\end{eqnarray}
where $f$ sums over all the (s)quarks and (s)leptons,
$N_c^f$ is the color factor, 3 for (s)quarks and 1 for (s)leptons,
and all the masses are implicitly $S_1,\ S_2$-dependent. 
$H$, $h$, $A$ and $G$ ($H^+$ and $G^+$)
label the neutral (charged) Higgs and Goldstone bosons,
$\tilde{\chi}^+_i$ and $\tilde{\chi}^0_i$ represent charginos and
neutralinos.
We have also used short-handed notations 
$\tilde f_i=m^2_{\tilde f_i}$, $W=m^2_W$, etc.
The function $G(x)$ is defined as
\begin{equation}
G(x)={x^2\over 32\pi^2}\biggl(\blog\ x - {3\over 2}\biggr),
\end{equation}
where $\blog\ x=\ln(x/Q^2)$, with $Q$ the renormalization scale.

\begin{figure}[thb]
\epsfysize=2in
\epsffile[40 320 350 500]{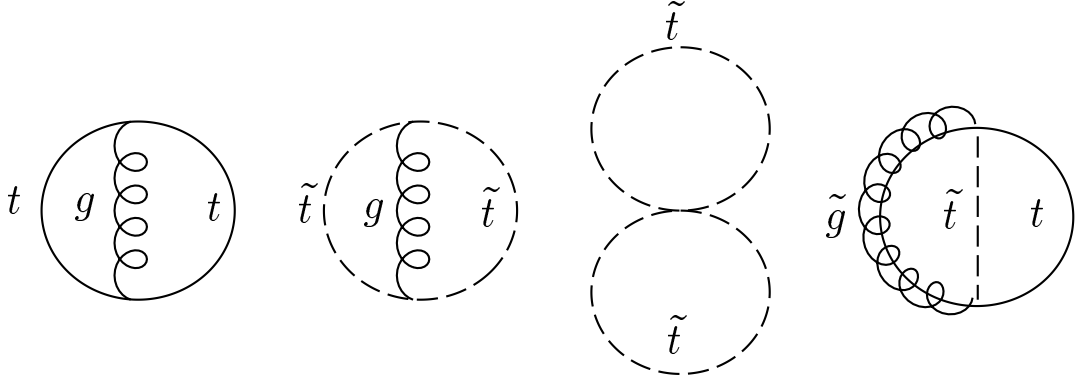}
\caption[]{Bubble diagrams for the two-loop effective potential
to the order of ${\cal O}(\lambda_t^2\alpha_s)$ in the MSSM. 
\label{feyn}}
\end{figure}

The two-loop effective potential can be derived similarly,
the corresponding Feynman diagrams are plot in Fig.~\ref{feyn}.
To the order of ${\cal O}(\lambda_t^2\alpha_s)$, we have
in the Landau gauge
\begin{eqnarray}
&&V_{\rm 2loop}(S_1,S_2)= 
32\pi\alpha_s \Biggl\{J(t,t)-2 t\ I(t,t,0) \nonumber\\
&& +{1\over 2}(c^4_t+s^4_t)\sum_{i=1}^2 J({\tilde t_i},{\tilde t_i}) 
+ 2 s^2_t c^2_t J({\tilde t_1},{\tilde t_2})
+ \sum_{i=1}^2 {\tilde t_i} I({\tilde t_i},{\tilde t_i},0)\nonumber\\
&& +\sum_{i=1}^2 L({\tilde t_i},{\tilde g},t) - 
4 m_{\tilde g}\ m_t\ s_t c_t
\biggl(I({\tilde t_1},{\tilde g},t)
-I({\tilde t_2},{\tilde g},t)\biggr)\Biggr\} 
\label{2lep}
\end{eqnarray}
where the (minimally) subtracted functions $I$ and $J$ are
defined in the Appendix, $s_t$ is the top-squark mixing angle.
This effective potential, in the limit of $\tan\beta\rightarrow\infty$
and no left-right squark-mixing ($s_t=0$), agrees with that of Ref.~\cite{HH}.
As a good  check, one can show  that the effective potential
$V(S_1,S_2)$ is invariant under the renormalization
scale change, up to 
two-loop terms which are ignored in our approximation.

The effective potential Eq.~(\ref{2lep}), as a generating functional,
encodes the information of two-loop tadpoles and self-energies
at zero external momentum. From this, one can find 
two-loop $CP$-even Higgs-boson masses by 
solving appropriate on-mass-shell conditions.

This proceeds as follows:
We first minimize $V(S_1,S_2)$ at the Higgs 
vacuum expectation values $S_1=v_1$ and $S_2=v_2$,
\begin{equation}
{\partial V\over\partial S_1}\biggr|_{S_1=v_1, S_2=v_2}\ =\ 0, \qquad
{\partial V\over\partial S_2}\biggr|_{S_1=v_1, S_2=v_2}\ =\ 0,
\end{equation}
which is equivalent to the requirement that
the tadpoles vanish,
\begin{eqnarray}
&& {T_1\over v_1}\ =\ {1\over2} m_Z^2c_{2\beta} + m_{H_1}^2
 + \mu^2 + B\mu\tan\beta\ ,\label{ewsb1}\\ 
&& {T_2\over v_2}\ =\ -{1\over2} m_Z^2 c_{2\beta} + m_{H_2}^2 + \mu^2 + 
B\mu\cot\beta\ ,\label{ewsb2}
\end{eqnarray}
where $\tan\beta=v_2/v_1,\ m^2_Z=(g^2+g'^2) v^2/4$ and $v^2=v^2_1+v^2_2$.
To the two-loop order,
the tadpoles $T_i, i=1,2$ are 
given by $T_i = T_i^{\rm 1loop}+ T_i^{\rm 2loop}$,
where the one- and two-loop tadpoles are defined by 
\begin{equation}
T_i^{\rm 1loop}\ =\ -\biggl({\partial V_{\rm 1loop}\over \partial S_i}\biggr)
\biggr|_{S_1=v_1,S_2=v_2}, \qquad 
T_i^{\rm 2loop}\ =\ -\biggl({\partial V_{\rm 2loop}\over \partial S_i}\biggr)
\biggr|_{S_1=v_1,S_2=v_2}.
\label{tad}\end{equation}
One can check that the one-loop tadpoles $T_i^{\rm 1loop}$
obtained in this way give the same results as in Ref.
\cite{BMPZ}\footnote{In Ref.~\cite{BMPZ}, we used
the 't Hooft-Feynman gauge. To compare with the one-loop tadpoles
obtained from Eq.~(\ref{1lep}), the Goldstone boson masses need to be 
set to zero.}, 
where they were explicitly calculated from the one-point function
Feynman diagrams (tadpole diagrams).

The $CP$-even Higgs-boson 
mass matrix, after some algebra, is 
\begin{eqnarray}
\label{massmatrix}
&&\qquad{\cal M}^2(p^2) \ =\  \\
&&\left(\begin{array}{cc}  m_Z^2 c_\beta^2 +
 m_A^2s_\beta^2 -{\rm Re}\Pi_{11}(p^2) +T_1/v_1 & -(
m_Z^2 +  m_A^2)s_\beta c_\beta -{\rm Re}\Pi_{12}(p^2)\\[2mm]
-( m_Z^2 +  m_A^2)s_\beta c_\beta -{\rm Re}\Pi_{12}(p^2)
&  m_Z^2s_\beta^2 +  m_A^2c_\beta^2 -{\rm Re}
\Pi_{22}(p^2) + T_2/v_2
\end{array}\right)\ ,\nonumber 
\end{eqnarray}
where $m_A^2=-\mu B(\tan\beta+\cot\beta)$, and both $m_Z$ and $m_A$ 
are $\dr$ running masses. $\Pi$'s represent two-point functions
(self-energies).

The radiatively corrected Higgs-boson
masses can be found by computing the zeroes of the inverse
propagator, $p^2 -{\cal M}^2(p^2)$. 
The complete one-loop formulae for self-energies 
at nonzero external momentum can be 
found, {\it e.g}, in Ref. \cite{BMPZ}.
For the two-loop self-energies, we shall approximate them 
as follows: 
\begin{equation}
\Pi_{ij}^{\rm 2loop}(0)\ =\ -\biggl({\partial^2 V_{\rm 2loop}
\over\partial S_i\partial S_j}\biggr)\biggr|_{S_1=v_1,
S_2=v_2}, \qquad i,j=1,2 . \label{self}
\end{equation}  
The difference of $\Pi_{ij}^{\rm 2loop}(0)$ and $\Pi_{ij}^{\rm 2loop}(p^2)$
is negligible.

\section{Numerical Procedure and Results}
 
We shall first show the improvement of renormalization scale
($Q$) dependence of the lightest $CP$-even Higgs-boson mass. 
We start by solving the two-loop renormalization group equations (RGEs) of the
MSSM. The boundary condition inputs for these RGEs are taken to be  
the observables $\alpha_s,\ \alpha_{em}$,  $G_F, \ M_Z,\ m_t,\ m_b$
and $m_\tau$ at the electro-weak scale, $M_Z$,  
and the universal scalar soft mass $M_0$, 
gaugino soft mass $M_{1/2}$ and
trilinear scalar coupling $A_0$
at the unification scale, $M_{\rm GUT}\approx 2 \times 10^{16}\ {\rm GeV}$
(This is sometimes called the minimal supergravity (mSUGRA) scenario.).
We have also properly taken into account the
low energy threshold corrections to the gauge and
Yukawa couplings, as in Ref. \cite{BMPZ}.

In this scenario, the $\mu$-parameter and the (tree-level) mass $m_A$ are
determined in terms of
other variables through the minimization conditions 
(\ref{ewsb1}) and (\ref{ewsb2}),
which ensure the correct electro-weak symmetry breaking, 
\begin{eqnarray}
&&\mu^2 = {1\over 2}\biggl[\tan 2\beta\biggl({\overline m}^2_{H_2}\tan\beta 
-{\overline m}^2_{H_1}\cot\beta\biggr)-m^2_Z\biggr] ,\\
&&m^2_A = {1\over \cos 2\beta}
\biggl({\overline m}^2_{H_2}-{\overline m}^2_{H_1}\biggr) - m^2_Z\ ,
\end{eqnarray}   
where ${\overline m}^2_{H_1}=m^2_{H_1}-T_1/v_1$ and
${\overline m}^2_{H_2}=m^2_{H_2}-T_2/v_2$.

We then calculate the lightest $CP$-even Higgs-boson mass from the
mass-squared matrix Eq.~(\ref{massmatrix}). We use the one-loop tadpole
and self-energy formulae from Ref.~\cite{BMPZ}, while for the
two-loop contributions, we compute the tadpoles and self-energies
numerically according to Eqs.~(\ref{tad}) and (\ref{self})
by replacing the differentiation by  a finite difference.
The field-dependent masses in Eq.~(\ref{2lep}) are the top-quark
mass $m_t = \lambda_t S_2/\sqrt{2}$ and the top-squark masses
which are found from the following field-dependent mass-squared matrix:
\begin{equation}
\left(\begin{array}{cc}
M^2_Q+{1\over2}\lambda_t^2S^2_2
	& {1\over\sqrt{2}}\lambda_t(A_tS_2+\mu S_1)\\[2mm]
{1\over\sqrt{2}}\lambda_t(A_tS_2+\mu S_1) 
	& M^2_U+{1\over2}\lambda_t^2S^2_2
\end{array}\right)\ ,
\end{equation}
where we have neglected the field-dependent $D$-term contributions
and $M_Q, M_U$ are squark soft masses at the low-energy scale. The 
field-dependent angle $s_t$ in Eq. (\ref{2lep}) is defined as the 
mixing angle of the above mass-squared matrix. 

\begin{figure}[thb]
\epsfysize=3.0in
\epsffile[30 240 300 525]{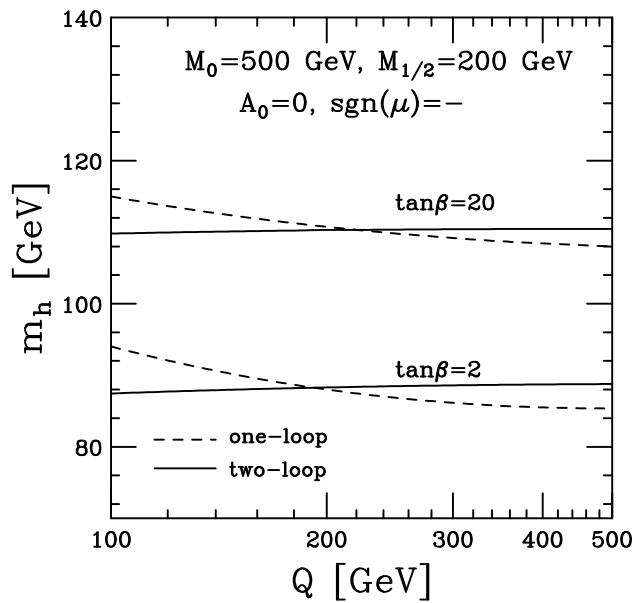}
\caption[]{Renormalization-scale ($Q$) 
dependence of the lightest $CP$-even Higgs-boson mass $m_h$.
The dashed and solid lines correspond to the one- and two-loop  
masses respectively. We have fixed the universal boundary conditions,
$M_0 = 500$ GeV, $M_{1/2}=200$ GeV, $A_0=0$, and chosen a negative
$\mu$-parameter.
\label{scale}}
\end{figure}

In Fig.~\ref{scale}, we show the dependence of 
one- and two-loop radiatively corrected $CP$-even 
Higgs-boson masses $m_h$ on the renormalization scale $Q$.
We choose the universal soft parameters $M_0 = 500$ GeV, 
$M_{1/2}=200$ GeV and $A_0=0$ at the unification scale, and set
the sign of $\mu$-parameter to be negative\footnote{Our 
convention of the sign of $\mu$-parameter 
is opposite to that of Refs.~\cite{CQW}
and \cite{hollik}.}.
We plot two choices of $\tan\beta$, 2 and $20$.
The one- and two-loop masses are shown in dashed and solid lines respectively.
In the corrections to the Higgs-boson masses $m_h$, we have used 
the $\dr$ running top-Yukawa coupling at the scale $Q$. 
The formulae which convert the top-quark pole mass $m_t^{\rm pole}$
to $\lambda_t(Q)$ are given in Ref.~\cite{BMPZ}.

We see that the one-loop radiatively corrected Higgs-boson masses
vary by about $10$ GeV as the renormalization scale $Q$
varies from $100$ to $500$ GeV. However, once we properly
include the two-loop radiative corrections, the scale dependence of
$m_h$ becomes much milder, and we find for all ranges of the scale
$m_h\simeq 88$ GeV and $110$ GeV for $\tan\beta=2$ and $20$ respectively.
The two-loop calculation can only change $m_h$ by a few GeV.

\begin{figure}[thb]
\epsfysize=3.0in
\epsffile[30 240 300 525]{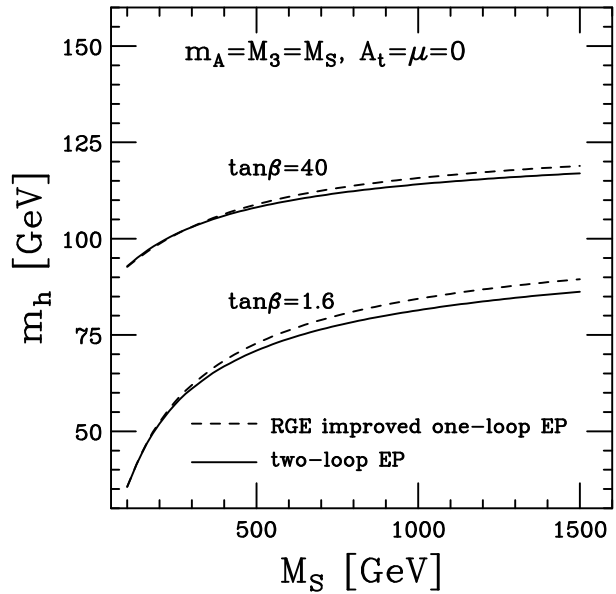}
\caption[]{Higgs boson masses $m_h$ vs. the squark soft masses 
$M_S$ in the no-squark-mixing case $A_t=\mu=0$. The solid lines are 
results from the two-loop EP approach. For comparison, we also
show the results from the RGE-improved one-loop EP approach
in dashed lines. 
\label{CQW1}}
\end{figure}

\begin{figure}[thb]
\epsfysize=3.0in
\epsffile[30 240 300 525]{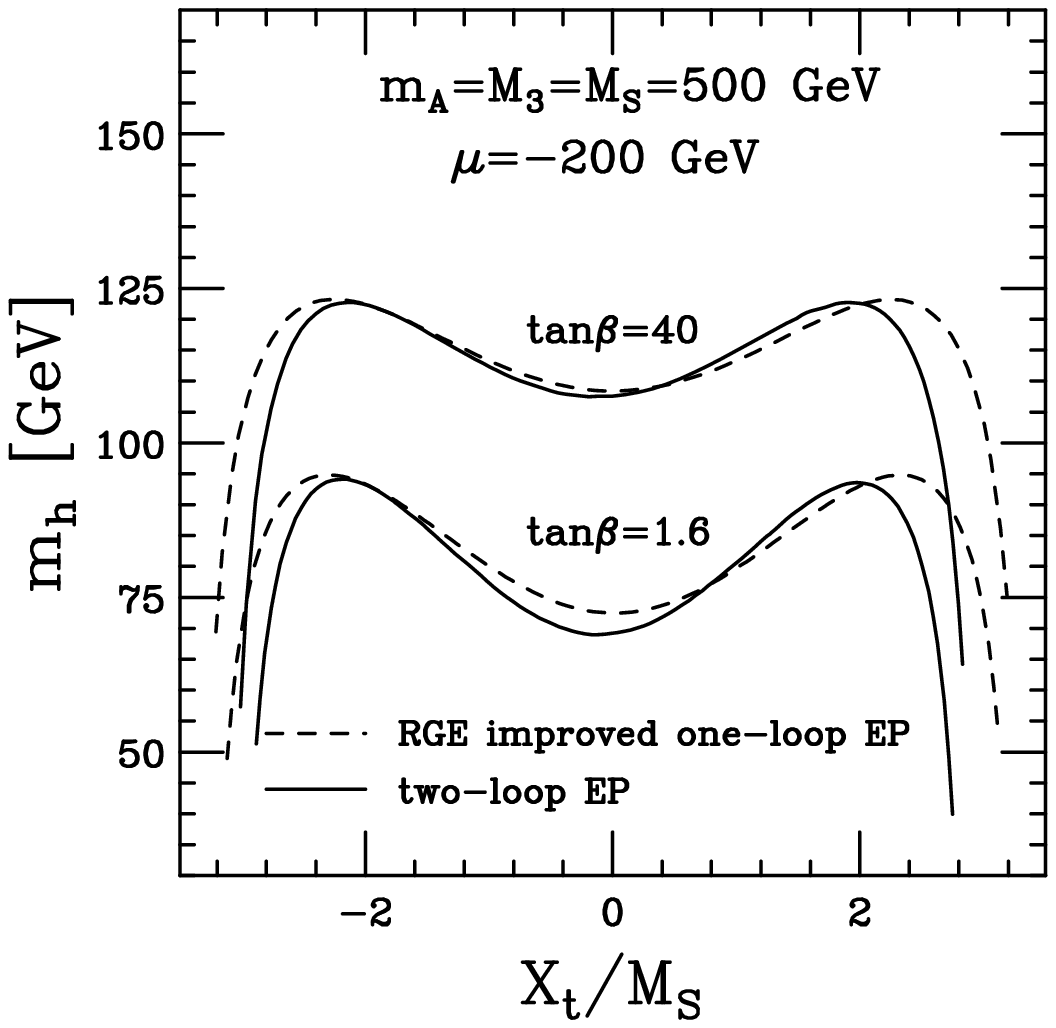}
\caption[]{Higgs-boson masses $m_h$ vs. $X_t/M_S$,
where $X_t=A_t+\mu/\tan\beta$. Both the results from
the RGE-improved one-loop EP and two-loop EP approaches
are shown.
\label{CQW2}}
\end{figure}

In Figs. \ref{CQW1} and \ref{CQW2}, we compare our results with
that of RGE-improved effective potential (EP) approach \cite{CQW}.  
In that approach, the heavy particles are decoupled at $m_{\tilde t}$ (or
$m_{\tilde t_1}$ and
$m_{\tilde t_2}$ stepwisely if they are very different). 
The two-loop RGEs of the effective field theory below the decoupling scale 
is then used to run the effective couplings to the
scale where the Higgs-boson mass is evaluated, {\it e.g.} 
the on-shell top-quark mass. Since the next-to-leading-order 
corrections are negligible at this scale \cite{Casas}, this allows an
analytical solution to the two-loop RGEs \cite{CQW}.

In this part of numerical analysis, we do not impose the
minimization conditions Eqs.~(\ref{ewsb1}) and (\ref{ewsb2}),
instead, the $\mu$-parameter and the $CP$-odd Higgs boson mass $m_A$
are taken as inputs. We further choose the squark soft masses
$M_Q=M_U=M_S$. In Fig.~\ref{CQW1}, we
show the two-loop Higgs-boson masses $m_h$ versus
$M_S$, in the RGE-improved one-loop EP approach
(dashed lines) and in the two-loop EP approach (solid lines). 
The parameters $A_t$ and $\mu$ are set to zero; this corresponds
to the no left-right squark-mixing limit. Other parameters are $m_A=M_3=M_S$.
Results from the two approaches generally agree to $\lsim 3$ GeV, 
for both small and large $\tan\beta$ cases.
The difference comes from the ${\cal O}(\lambda_t^4)$ corrections
which are neglected in this calculation \cite{hollik}.

In Fig. \ref{CQW2}, we choose a nonzero $\mu$-parameter, 
$\mu=-200$ GeV, and plot $m_h$ versus $X_t/M_S$, where 
$X_t=A_t+\mu/\tan\beta$ is related to the off-diagonal element
of the squark-mass matrix. We see the results from the
two approaches still agree quite well except for the
large $X_t/M_S$. In particular, the curves
for the two-loop EP approach peak at $X_t/M_S\simeq\pm 2$, which 
is different from the RGE-improved one-loop EP approach. 
This however agrees with the
results of a recent analysis \cite{hollik}. 
The slight asymmetry of the solid curves with respect to the
reflection, $X_t\rightarrow -X_t$, originates from the last term in 
Eq. (\ref{2lep}).
We found that for
$\tan\beta=40$ the upper limit
for the Higgs-boson mass is $125$ GeV 
at the  large squark-mixing ($X_t/M_S=\pm 2$).

\section{Conclusions}

To conclude, we have used an effective potential
method to calculate the two-loop corrections to the 
lightest $CP$-even Higgs-boson mass in the MSSM. Our approach 
is straightforward and easy to program, and can be extended to include
two-loop corrections of order ${\cal O}(\lambda_t^4)$. 
We show that the renormalization scale dependence of 
$m_h$ improves after including the two-loop corrections,
this largely reduces the uncertainty associated with one-loop calculations.
We have shown that the two-loop correction 
is only about a few GeV with respect 
to the one-loop results (where the $\dr$ running coupling
$\lambda_t$ is used).
We have also compared our results with some previous 
two-loop calculations. We found good agreements with
the RGE-improved one-loop EP approach of Ref.~\cite{CQW}
except for the case of large left-right 
squark mixing, where we obtained similar results
as in Ref.~\cite{hollik}. The upper bound for the Higgs-boson
mass $m_h\lsim 125$ GeV is achieved at the region of parameter 
space for large $\tan\beta$ and left-right
squark mixings.

\vspace{.7cm}
{\noindent\Large\bf Acknowledgements}
\vspace{.7cm}

I would like to thank K. Matchev for participating in the early
stage of this work,  J. Bagger, T. Han and C. Kao for conversations and
comments, and  C. Wagner and G. Weiglein
for communications and numerical comparisons. 
This work was supported in part by 
a DOE grant No. DE-FG02-95ER40896
and in part by the Wisconsin Alumni Research Foundation.

\vspace{.7cm}
{\noindent\Large\bf Appendix: The functions $I(x,y,z)$ and $J(x,y)$}
\vspace{.7cm}
\setcounter{equation}{0}
\renewcommand{\theequation}{A.\arabic{equation}}

Momentum integrals arising from the two-loop bubble diagrams 
have one-loop subdivergences which can 
be subtracted in the standard way \cite{FJJ}. Furthermore, in the
$\dr$ scheme there is no complication associated with vector-boson
loops\footnote{In contrast,
the vector bosons live
in $d=4-2\epsilon$ dimensions
in the \mbox{\footnotesize$\overline{\rm MS}$~} scheme, 
in reducing the two-loop integrals to 
the form of $I$ and $J$, there will be extra finite terms
from themselves and the associated one-loop subdiagrams.} ,
so all integrals can be expressed in terms of (minimally) subtracted functions 
$I,\ J$  which
are \cite{FJJ}
\begin{equation}
(16\pi^2)^2 J(x,y)\ =\ x\ y\biggl[1-\blog\ x
-\blog\ y+\blog x\ \blog\ y\biggr],
\end{equation}
\begin{eqnarray}
&&(16\pi^2)^2 I(x,y,z)\ =\ -{1\over 2}\biggl[(y+z-x)\blog\ y\ \blog\ z
+(z+x-y)\blog\ z\ \blog\ x\\
&&+(x+y-z)\blog\ x\ \blog\ y -4(x\ \blog\ x 
+  y\ \blog\ y+ z\ \blog\ z)+\xi(x,y,z)+5(x+y+z)\biggr]\nonumber,
\end{eqnarray}
where $\xi$ is given by
\begin{equation}
\xi(x,y,z)\ =\ 8b\biggl[L(\theta_x)+L(\theta_y)+L(\theta_z)-{\pi\over2}\ln 2
\biggr]
\end{equation}
when $-b^2=a^2=(x^2+y^2+z^2-2xy-2xz-2yz)/4\leq 0$,
and 
\begin{equation}
\xi(x,y,z)\ =\ 8a\biggl[-M(-\phi_x)+M(\phi_y)+M(\phi_z)\biggr]
\end{equation}
when $a^2>0$. Here $L(t)$ is Lobachevsky's function, defined as
\begin{equation}
L(t)\ =\ -\int_0^t dx \ln \cos\ x\ =\ t\ln 2 
-{1\over2}\sum_{k=1}^\infty (-)^{k-1}{\sin 2kt\over k^2},
\end{equation}
and the function $M(t)$ is defined as
\begin{equation}
M(t)\ =\ -\int_0^t dx \ln \sinh\ x\ =\ {\pi^2\over12} - {t^2\over2}
+t\ln 2 - {1\over2}{\rm Li}_2(e^{-2t}),
\end{equation}
${\rm Li}_2$ is the dilogarithm function.

The angles $\theta_{x,y,z}$ and $\phi_{x,y,z}$ are defined by
\begin{equation}
\theta_x\ =\ \arctan\biggl({y+z-x\over 2b}\biggr),
\qquad \phi_x\ =\ {\rm arccoth}\biggl({y+z-x\over 2a}\biggr), \qquad {\rm etc.}
\end{equation}

Finally we have also used 
the following function in the two-loop effective
potential:
\begin{equation}
L(x,y,z) = J(y,z)-J(x,y)-J(x,z)-(x-y-z)I(x,y,z).
\end{equation}

\end{document}